\documentclass{article}
\usepackage{spconf,amsmath,graphicx}
\usepackage{cite}
\usepackage{amssymb,amsfonts,amsthm,bm}
\usepackage[usenames,dvipsnames]{xcolor}

\usepackage{booktabs}
\usepackage{graphicx}
\usepackage{multirow}
\usepackage{stfloats}
\usepackage{array} 
\usepackage{caption}

\definecolor{mygreen}{rgb}{0,0.5,0}
\definecolor{darkblue}{RGB}{0,0,150}
\title{A Deep Representation Learning-based Speech Enhancement Method Using Complex Convolution Recurrent Variational Autoencoder}
\name{Yang Xiang, Jingguang Tian, Xinhui Hu, Xinkang Xu, ZhaoHui Yin} 
\address{Hithink RoyalFlush AI Research Institute, Zhejiang, China \\
\{xiangyang2, tianjingguang, huxinhui, xuxinkang, yinzhaohui2\}@myhexin.com}
\vspace{-0.5cm}
\begin{document}
\ninept
\maketitle
\ninept
\vspace{-0.2cm}
\begin{abstract}
Generally, the performance of deep neural networks (DNNs) heavily depends on the quality of data representation learning. Our preliminary work has emphasized the significance of deep representation learning (DRL) in the context of speech enhancement (SE) applications. Specifically, our initial SE algorithm employed a gated recurrent unit variational autoencoder (VAE) with a Gaussian distribution to enhance the performance of certain existing SE systems. Building upon our preliminary framework, this paper introduces a novel approach for SE using deep complex convolutional recurrent networks with a VAE (DCCRN-VAE). DCCRN-VAE assumes that the latent variables of signals follow complex Gaussian distributions that are modeled by DCCRN, as these distributions can better capture the behaviors of complex signals. Additionally, we propose the application of a residual loss in DCCRN-VAE to further improve the quality of the enhanced speech. {Compared to our preliminary work, DCCRN-VAE introduces a more sophisticated DCCRN structure and probability distribution for DRL. Furthermore, in comparison to DCCRN, DCCRN-VAE employs a more advanced DRL strategy. The experimental results demonstrate that the proposed SE algorithm outperforms both our preliminary SE framework and the state-of-the-art DCCRN SE method in terms of scale-invariant signal-to-distortion ratio, speech quality, and speech intelligibility.}
\end{abstract}

\section{Introduction}
\label{sec:intro}
The aim of speech enhancement (SE) \cite{loizou2013speech} is to remove background noise and improve the quality and intelligibility of the observed speech. Recently, driven by the growing demand for online meeting systems, SE is required to reduce the word error rate for accurate live captioning when transmitting high-quality speech audio in complex noise conditions \cite{eskimez2021human, iwamoto2022bad}. Thus, SE has become a hot research topic.

Over the past few decades, many single-channel speech enhancement (SE) algorithms have been developed. Classical SE algorithms include the Wiener filtering-based strategies \cite{lim1978all}, signal subspace methods \cite{christensen2016experimental}, codebook-based approaches \cite{srinivasan2005codebook}, and non-negative matrix factorization (NMF) techniques \cite{mohammadiha2013supervised, xiang2020nmf, xiang2021novel}. In recent times, driven by advancements in hardware and deep neural network (DNN) techniques, DNNs have become widely utilized in SE tasks \cite{ wang2018supervised, sun2017multiple, luo2019conv, xiang2020parallel, hu20g_interspeech, wang2021compensation}. These DNN-based methods often adopt diverse DNN architectures \cite{sun2017multiple} to predict various targets (e.g., different masks \cite{wang2014training}, speech spectrum \cite{xu2013experimental}, waveform \cite{luo2019conv}, and speech presence probability \cite{tu2019speech}) for SE. Unlike classical SE algorithms, DNN-based methods \cite{boll1979suppression, lim1978all, christensen2016experimental, srinivasan2005codebook} typically rely on fewer statistical assumptions \cite{xu2014regression}, and can employ nonlinear processes to model complex high-dimensional signals. Consequently, DNNs tend to offer a more practical and reasonable solution for various applications.

While many DNN-based SE methods have been developed \cite{wang2018supervised}, the majority of these methods overlook the significance of reliable representations for DNN and give less consideration to the application of DNNs for obtaining better speech representations during SE. Consequently, these algorithms may not consistently perform well in environments with complex noise. Additionally, this oversight contributes to many SE algorithms that aim to enhance human listening but inadvertently degrade ASR system performance due to inexplicable signal distortions \cite{eskimez2021human, iwamoto2022bad}. Acquiring reliable representations is highly significant for DNN-based SE algorithms because one of the fundamental principles of DNNs is that DNNs are built upon data representation learning \cite{chan2021white, bengio2013representation, dai2021closed} to generate relevant data. A strong signal representation not only enhances DNNs' capacity to extract valuable information in complex environments \cite{xie2021disentangled, bengio2013representation}, but it also contributes to an improved DNN's predictive ability \cite{bengio2013representation}. Moreover, a solid representation can reduce the demand on learning machines, facilitating successful task execution \cite{wang2018supervised}. 

Due to the importance of representation for DNN \cite{bengio2013representation, dai2021closed}, recent research endeavors have explored deep representation learning (DRL)-based SE algorithms \cite{carbajal2021guided, fang2021variational, carbajal2021disentanglement}. Broadly, these methods leverage variational autoencoders (VAEs) \cite{kingma2013auto} to acquire speech representations and enhance the generalization capacity of these algorithms. However, these VAE-based approaches often struggle to disentangle speech representations from the underlying noise, resulting in inaccurately acquired speech representations and potentially suboptimal SE performance \cite{bando2018statistical}. To address this problem, our preliminary work \cite{xiang2022bayesian} introduced a novel VAE SE framework named  PVAE. PVAE applies a conditional posterior assumption to derive a novel evidence lower bound (ELBO), enabling VAEs to effectively disentangle distinct signal representations for superior speech representation. Empirical experiments \cite{xiang2022bayesian} indicated that our DRL strategy can enhance the performance of traditional DNN-based SE methods \cite{huang2014deep}. Building upon the prior work \cite{xiang2022bayesian}, we proposed the integration of $\beta$-VAE \cite{higgins2016beta, xiang2022deep} and adversarial training (GANs) \cite{goodfellow2014generative, xiang2022two} to further enhance PVAE's representation learning and signal generation capabilities, respectively, thus achieving better SE performance.

While our preliminary SE work of disentanglement VAE (DIS-VAE) \cite{xiang2022two} has confirmed the effectiveness of DRL in DNN-based SE algorithms, our DRL strategy falls short of achieving state-of-the-art (SOTA) SE performance. This limitation stems from employing a basic gated recurrent unit (GRU) structure and relying on a simplistic posterior assumption (multivariate normal distributions) for SE. Additionally, our use of the log-power spectrum (LPS) as the training target excludes phase information, which hampers SE performance. To enhance DIS-VAE's SE performance, this study modifies the deep complex convolutional recurrent networks (DCCRN) \cite{hu20g_interspeech}  and integrates them into DIS-VAE \cite{xiang2022two}. DCCRN, a SOTA SE architecture, combines a convolutional recurrent network (CRN) with a convolutional encoder-decoder (CED) structure and long short-term memory (LSTM). This incorporation of DCCRN offers two key advantages for DIS-VAE \cite{xiang2022two}. Firstly, CRN extracts high-level features from input data, while LSTM effectively models temporal dependencies, resulting in improvements over our previous GRU-based structure. Secondly, DCCRN can adeptly handle complex-value operations, avoiding the loss of phase information and making it well-suited for time-frequency (T-F) complex spectrum analysis of signals. {On the other hand, DCCRN, while being a SOTA SE method, overlooks the importance of DRL in a DNN structure. DCCRN directly estimates a complex mask for SE and does not consider how to obtain a more reliable representation and enhance the decoder's generative ability based on the acquired representations. This oversight may hinder DCCRN from achieving better SE performance \cite{chan2021white, bengio2013representation, wang2018supervised}. However, this limitation can be mitigated by our preliminary DRL strategy \cite{xiang2022two}. To better align our DRL strategy \cite{xiang2022two} with the DCCRN structure, we incorporate the complex VAE \cite{nakashika2020complex} into our preliminary system \cite{xiang2022two}, and propose a novel SE system, named DCCRN-VAE. Specifically, we assume that latent variables for speech and noise follow complex Gaussian distributions (CGD). The disentanglement of signals \cite{xiang2022two} is conducted based on CGD, enabling us to capture complex signal behaviors more accurately than the previous Gaussian distribution-based approaches \cite{xiang2022two}. Moreover, we employ residual loss \cite{he2016deep}  and adversarial training \cite{goodfellow2014generative} to enhance the DRL and generative capability of DCCRN-VAE, respectively. Compared to DIS-VAE \cite{xiang2022two}, the novel DCCRN-VAE SE system employs a more sophisticated probabilistic model and DNN framework for signal modeling. DCCRN-VAE also improves the DCCRN's DRL ability. Furthermore, in this work, incorporating a large-scale dataset (500 hours) enhances the practicality of our model.}

{\section{Proposed DCCRN-VAE SE system}
\vspace{-0.2cm}
\label{sec:Complex}
This section will first introduce the signal model of our system. Afterward, we will demonstrate how we modify the DCCRN to adapt it to our novel DCCRN-VAE strategy. Finally, we will explain the details of the proposed DCCRN-VAE.}

\begin{figure*}[!tbp]
  \centering
  \setlength{\abovecaptionskip}{0.1cm}
  \centerline{\includegraphics[scale=0.45]{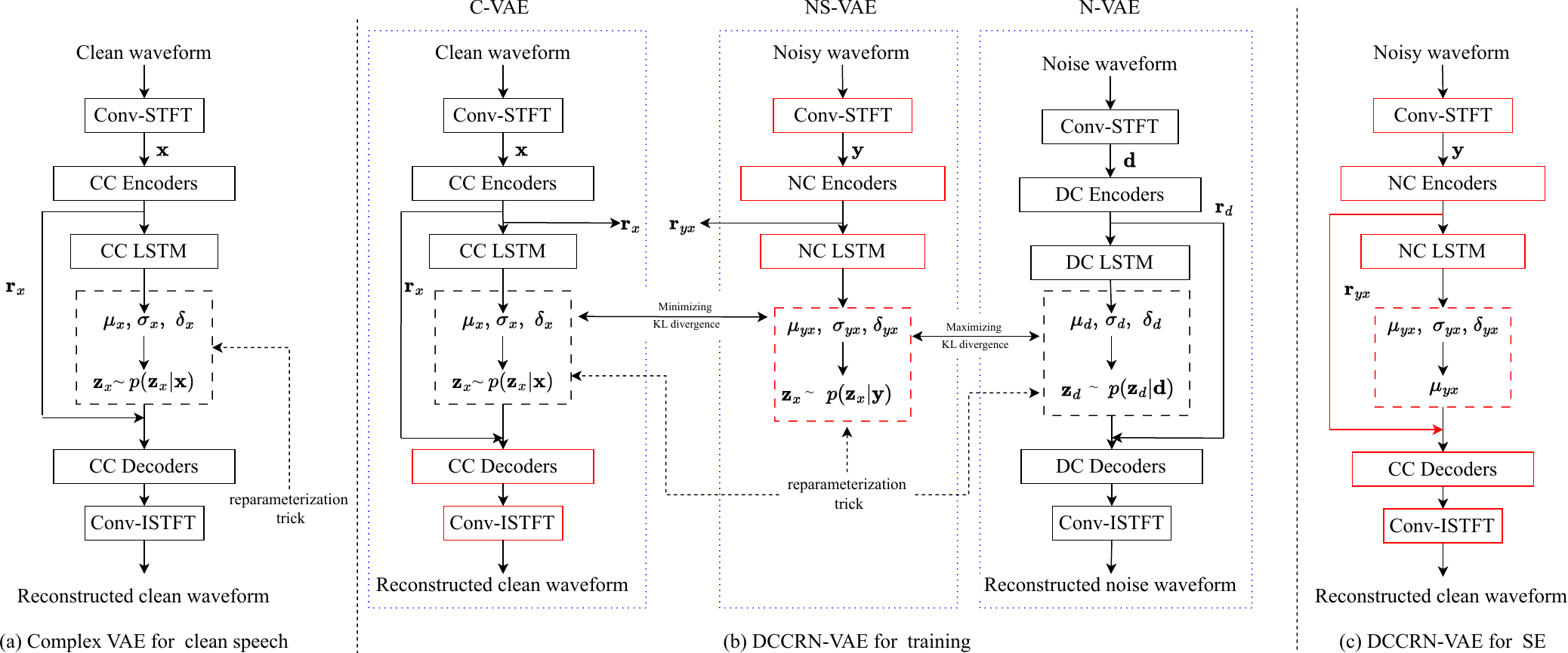}}
  \setlength{\belowcaptionskip}{3pt}
  \caption{Graphical illustration for DCCRN-VAE.}
  \label{fig:DCCRN_VAE}
  \vspace{-0.7cm}
\end{figure*}

\vspace{-0.2cm}
\subsection{Signal Model}
In a noisy environment, the observed signal is \(y(t)\), where \(t\) is the time index. The purpose of SE is to extract clean speech \(x(t)\) and remove noise \(d(t)\) from the observed signal \(y(t)\). Utilizing the short-time Fourier transform (STFT), the complex spectrum vector of the observed signal, clean speech, and noise at each frame can be denoted \(\mathbf{y} \in \mathbb{C}^F\), \(\mathbf{x} \in \mathbb{C}^F\), and \(\mathbf{d} \in \mathbb{C}^F\), respectively. \(F\) represents the number of frequency bins. In this work, we assume that \(\mathbf{y}\) can be generated from a random process involving speech latent representation variables \(\mathbf{z}_x \in \mathbb{C}^L\) and noise latent representation variables \(\mathbf{z}_d \in \mathbb{C}^L\) (\(L\) is the dimension of latent variables). Similarly, \(\mathbf{x}\) and \(\mathbf{d}\) can be independently generated using \(\mathbf{z}_x\) and \(\mathbf{z}_d\), respectively. Additionally, we assume that \(\mathbf{z}_x\) and \(\mathbf{z}_d\) can be estimated from the speech and noise posterior distributions \(p(\mathbf{z}_x|\mathbf{x})\) and \(p(\mathbf{z}_d|\mathbf{d})\), respectively, or from the posterior distributions of noisy speech, \(p(\mathbf{z}_x|\mathbf{y})\) and \(p(\mathbf{z}_d|\mathbf{y})\) \cite{xiang2022bayesian}. To perform SE, it is necessary to disentangle the different latent variables from the observed signal.

\vspace{-0.2cm}
\subsection{Modified Complex VAE}
The general VAE \cite{kingma2013auto} establishes a probabilistic generative process that connects clean speech with its latent variables. It introduces a principled approach to simultaneously learn latent variables, generative models, and recognition models. The models training involves maximizing the ELBO \cite{kingma2013auto}
\begin{equation}
\footnotesize
 \begin{aligned}
   &  {\mathbb E_{{\bf {x}} \sim p(\bf {x})}[\log q(\bf {x})]} \ge -{\mathcal{L}}_c, \\
   & {\mathcal{L}}_c = {\mathbb E_{{\bf {x}} \sim p(\bf {x})}} \left[D_{KL}\left({p({\bf z}_x|{\bf{x}}))}||{q({{\bf z}_x})}\right)\right] \\
   & \quad \quad - {\mathbb E_{{\bf {x}} \sim p(\bf {x})}} \left[ {\mathbb E_{{{\bf {z}}_x} \sim p({{\bf {z}}_x}|\bf {x})}}\left[\log {q({\bf{x}}|{\bf z}_x)} \right]\right],
   \end{aligned}
  \label{ELBO}
\end{equation}
where $D_{KL}(||)$ denotes the KL divergence. Maximizing this lower bound is equivalent to minimizing ${\mathcal{L}}_c$.

The complex VAE \cite{nakashika2020complex} shares similarities with the conventional VAE \cite{kingma2013auto}, both optimizing the same final objective (\ref{ELBO}). The key distinction lies in the utilization of a complex-valued DNN for modeling complex data. Specifically, the complex VAE employs CGD \cite{nakashika2020complex}, estimated by a complex DNN, to define prior and posterior distributions of signals. In contrast to the general VAE \cite{kingma2013auto, xiang2022bayesian}, the complex VAE employs a more sophisticated probabilistic model and DNN framework for analyzing speech signal spectrum. Notably, it addresses the problem of preserving phase information \cite{nakashika2020complex}, an aspect often overlooked in signal spectrum analysis due to the unpredictable nature of phase. This distinctive feature endows complex VAE with a robust spectrum modeling capability.

To compute (\ref{ELBO}), it becomes essential to establish the corresponding posterior and prior distributions and predefine \(q({\bf z}_x)\). In line with \cite{nakashika2020complex}, we posit that \(p({\bf z}_x|{\bf{x}})\) is a complex normal distribution with diagonal covariance and pseudo-covariance matrices
\begin{equation}
\footnotesize
   \begin{aligned}
   & {p({\bf z}_x|{\bf{x}})} = {\mathcal{N}}_c \left({{\bf {{\bf z}}}_x;{\bf \mu}_{\theta_{x}}({\bf{x}}),{\sigma}_{\theta_{x}}({\bf x})\bf{I}}, {\bf{\delta}}_{\theta_{x}}({\bf{x}})\bf{I}\right),
  \end{aligned}
  \label{posterior_c}
\end{equation}
where $\mathbf{I}$ is the identity matrix; ${\mu}_{\theta_{x}}(\mathbf{x})$, ${\sigma}_{\theta_{x}}(\mathbf{x})$, and ${\delta}_{\theta_{x}}(\mathbf{x})$ can be estimated by the complex VAE's encoder $G_{\theta_x}(\mathbf{x})$ using the complex parameter $\theta_x$ (as shown in Fig.~\ref{fig:DCCRN_VAE}(a)). Here, ${\mu_x \in \mathbb{C}^L}$, ${\sigma_x \in \mathbb{R}^L_+}$, and ${\delta_x \in \mathbb{C}^L}$ represent the mean, covariance, and pseudo-covariance in the complex Gaussian distribution, respectively. The distribution $q({\mathbf{z}_x})$ is predefined as a standard complex normal distribution
\begin{equation}
\footnotesize
   \begin{aligned}
   & q({\bf z}_x) = \mathcal{N}_c ({{\bf {{\bf z}}}_x;{\bf{0}},{\bf{I}}, {\bf{0}}}).
  \end{aligned}
  \label{prior_single}
\end{equation}
Based on the VAE theory \cite{kingma2013auto, higgins2016beta}, the second term $\log {q({\mathbf{x}}|\mathbf{z}_x)}$ in (\ref{ELBO}) is responsible for signal reconstruction. In \cite{nakashika2020complex}, ${q({\mathbf{x}}|\mathbf{z}_x)}$ is assumed to follow the CGD with unit covariance and zero pseudo-covariance. {To adapt to the SE task and improve speech quality and intelligibility, this work applies the signal-to-distortion ratio (SI-SDR) \cite{le2019sdr} loss to replace the second term in (\ref{ELBO}) for signal recovery. Thus, we have}
\begin{equation}
  \footnotesize
   \begin{aligned}
   & {\mathbb E_{{\bf {x}} \sim p(\bf {x})}} \left[ {\mathbb E_{{{\bf {z}}_x} \sim p({{\bf {z}}_x}|\bf {x})}}\left[\log {q({\bf{x}}|{\bf z}_x)} \right]\right] \approx \text { SI-SNR }({\bf x}_t, \hat{\bf x}_t) \\
 & \quad \quad \quad \text {SI-SNR}({\bf x}_t, \hat{\bf x}_t) =10 \log _{10} \frac{\left\|\mathbf{x}_{\text {target}}\right\|^2}{\left\| \hat{\mathbf{x}}_t-\mathbf{x}_{\text {target }}  \right\|^2} \\
 & \quad \quad \quad \quad \quad \quad \mathbf{x}_{\text {target }}=\frac{\langle\hat{\mathbf{x}}_t, \mathbf{x}_t\rangle \mathbf{x}_t}{\|\mathbf{x}_t\|^2},\\
  \end{aligned}
  \label{si_sdr_loss}
\end{equation}
where $\mathbf{x}_t$ and $\mathbf{\hat x}_t$  represent the original and reconstructed clean speech waveform vector, respectively. $\mathbf{\hat x}_t$ can be obtained through the complex VAE's decoder $G_{\varphi_x}(\mathbf{z}_x)$ with parameter $\varphi_x$.

In this work, the structure of our complex VAE is designed based on DCCRN \cite{hu20g_interspeech}. The modified DCCRN structure for complex VAE, as illustrated in Fig.~\ref{fig:DCCRN_VAE}(a), comprises several key components: Conv-STFT, Conv-ISTFT, complex clean encoders (CC Encoders), a representation module, complex clean LSTM (CC LSTM), and complex clean decoders (CC Decoders). The Conv-STFT and Conv-ISTFT components utilize 1D convolutions for STFT and inverse STFT operations. CC Encoders consist of a complex 2-D convolutional neural network (Conv2d), complex batch normalization, and real-valued Parametric Rectified Linear Unit (PReLU), consistent with the original DCCRN \cite{hu20g_interspeech}. The CC LSTM produces three outputs: the mean, covariance, and pseudo-covariance parameters of the complex Gaussian distribution. Following this, a representation module is employed, utilizing a complex Gaussian representation trick \cite{nakashika2020complex} to enable back-propagation from decoder to encoder. Finally, CC Decoders are responsible for signal recovery. Notably, including skip-connections (SC) \cite{he2016deep} aids in gradient flow by connecting the encoder and decoder, a critical factor in generating high-quality speech. We define the residual input for CC Decoders, derived from CC Encoders, as ${\bf r}_x$.

\vspace{-0.2cm}
\subsection{DCCRN-VAE for SE}
\label{sec:DCCRN_VAE}

To obtain a better SE performance \cite{xiang2022two}, we introduce a DCCRN-VAE SE system in this work, as illustrated in Fig.~\ref{fig:DCCRN_VAE}. DCCRN-VAE comprises a DCCRN-based clean speech VAE (C-VAE), noise VAE (N-VAE), and a noisy speech VAE encoder (NS-VAE). The objectives of C-VAE and N-VAE are to obtain fully disentangled representations of clean speech and noise, respectively. C-VAE and N-VAE have the same DCCRN structure, as shown in Fig.\ref{fig:DCCRN_VAE}(a), (b). The DC encoders, DC LSTM, and DC decoders correspond to the complex noise encoders, LSTM, and decoders, respectively. The objective of NS-VAE is to disentangle speech representation from the observed noisy speech. To achieve this, NS-VAE's encoder is used to estimate the posterior $p({\mathbf{z}_x}|{\mathbf{y}})$ with a  CGD
\begin{equation}
\footnotesize
   \begin{aligned}
   & {p({\bf z}_x|{\bf{y}})} = {\mathcal{N}}_c \left({{\bf {{\bf z}}}_x;{\bf \mu}_{\theta_{yx}}({\bf{y}}),{\sigma}_{\theta_{yx}}({\bf y})\bf{I}}, {\bf{\delta}}_{\theta_{yx}}({\bf{y}})\bf{I}\right),
  \end{aligned}
  \label{posterior_noi}
\end{equation}
where ${\mu}_{\theta_{yx}}(\mathbf{y})$, ${\sigma}_{\theta_{yx}}(\mathbf{y})$, and ${\delta}_{\theta_{yx}}(\mathbf{y})$ can be estimated by the complex VAE's encoder $G_{\theta_{yx}}(\mathbf{y})$ using the complex parameter $\theta_{yx}$. In Fig.~\ref{fig:DCCRN_VAE}(b), NC encoders and LSTM represent the noisy complex encoders and LSTM, respectively.  Here, ${\mu_{yx}}$, ${\sigma_{yx}}$, and ${\delta_{yx}}$ represent the CGD's mean, covariance, and pseudo-covariance, respectively.

In DIS-VAE \cite{xiang2022deep, xiang2022two}, we leveraged the properties of $\beta$-VAE and DRL to derive an ELBO for signal disentanglement in the latent space, which indicates that SE performance relies heavily on the quality of learned clean speech representations \cite{xiang2022deep}. {However, DIS-VAE \cite{xiang2022two} only utilized a basic GRU and Gaussian distribution to validate the algorithm's correctness in improving representations. This work incorporates the DCCRN architecture and CGD to enhance SE performance to obtain superior representations. Additionally, compared to DCCRN, DCCRN-VAE employs a more sophisticated representation learning approach.} Building upon the principles of DCCRN \cite{hu20g_interspeech} and DIS-VAE \cite{xiang2022two}, we propose to leverage KL loss, residual loss, and adversarial loss for the DCCRN-VAE SE system.

The KL and residual losses aim to facilitate the representation learning. {The signal separation process is mainly based on them \cite{xiang2022two}}. In DIS-VAE, the KL loss was employed to disentangle speech and noise latent representations. Building upon this concept, to obtain improved clean speech representations, we propose minimizing the KL divergence (KLD) between $p({\mathbf{z}_x}|{\mathbf{y}})$ and $p({\mathbf{z}_x}|{\mathbf{x}})$, while maximizing the KLD between $p({\mathbf{z}_x}|{\mathbf{y}})$ and $p({\mathbf{z}_d}|{\mathbf{d}})$ because we assume that speech and noise posterior are independent \cite{xiang2022two}. There is a noteworthy difference from our previous work \cite{xiang2022two}. In the current setup, NS-VAE does not estimate $p({\mathbf{z}_d}|{\mathbf{y}})$ due to the SC's presence in the DCCRN framework. Without SC, DCCRN cannot generate high-quality speech \cite{he2016deep}. Nevertheless, in real SE environments, it is impossible to obtain the clean residual ${\bf r}_x$, necessitating us to ensure that the residual ${\bf r}_{yx}$ from NS-VAE's encoder closely approximates ${\bf r}_x$. To achieve a more accurate estimation of ${\bf r}_x$, we do not use NS-VAE to estimate $p({\mathbf{z}_d}|{\mathbf{y}})$ and instead define the residual loss. Thus, the KL loss $\mathcal{L}_{kl}$ and the residual loss $\mathcal{L}_{re}$ are expressed as
\begin{equation}
\footnotesize
 \begin{aligned}
  & \mathcal{L}_{kl} = {\mathbb E_{{\bf {y}} \sim p(\bf {y}),{\bf {x}} \sim p(\bf {x})}} \left[D_{KL}\left({p({\bf z}_x|{\bf{y}})}||{p({\bf z}_x|{\bf{x}})}\right)\right] \\
   & \quad \quad - \alpha{\mathbb E_{{\bf {y}} \sim p({\bf {y}}), {\bf {d}} \sim p({\bf {d}})}}  \left[D_{KL}\left({p({\bf z}_d|{\bf{y}})}||{p({\bf z}_d|{\bf{d}})}\right)\right]\\
  & \mathcal{L}_{re} = ||{\bf r}_{x}-{\bf r}_{yx}||_2^2, \quad  \mathcal{L}_{la} = \mathcal{L}_{kl} + \mathcal{L}_{re}, 
   \end{aligned}
  \label{latent_loss}
\end{equation}
where $\mathcal{L}_{la}$ is the latent representation loss, and $\alpha$ is an adjustable hyperparameter. Due to KLD without an upper bound, maximizing it may lead to DNN convergence issues. To address this, we assign a small weight to the noise KLD term, precisely $\alpha = 0.25$. Here, we use the sampling method to compute the KLD. {Additionally, $\mathcal{L}_{la}$ also ensures that DCCRN has a better DRL ability \cite{xiang2022two}.}

Following our previous work \cite{xiang2022two}, the adversarial loss is employed to enhance the generative capability of the CC decoders, taking into account the estimated clean speech representation ${\bf z}_x \sim p({\bf z}_x|{\bf{y}})$. {The main role of the adversarial loss (\ref{GAN_dc}) is to convert low-dimensional representations back to high-dimensional signals, focusing on signal reconstruction \cite{xiang2022two}, which is not considered in DCCRN. Thus, it also improves DCCRN's generative ability.} By introducing a discriminator $D_{\theta{dx}}(\cdot)$ with parameters $\theta_{dx}$, in accordance with the adversarial loss used in \cite{xiang2022two}, the adversarial loss is
\begin{equation}
\footnotesize
 \begin{aligned}
  & \mathcal{L}_{gan_c}(G_{\varphi_x}) = \mathbb{E}_{{{\bf {z}}_x} \sim p({{\bf {z}}_x}|{\bf {y}})}[(D_{\theta_{dx}}(G_{\varphi_x}({\mathbf{z}}_{x}))-1)^2] \\
  & \quad - \text {SI-SNR}({\bf x}_t, {\mathbb{E}_{{{\bf {z}}_x} \sim p({{\bf {z}}_x}|{\bf {y}})}}(G_{\varphi_x}({\mathbf{z}}_{x}))) \\
  & \mathcal{L}_{gan_c}(D_{\theta_{dx}}) = \mathbb{E}_{{{\bf {z}}_x} \sim p({{\bf {z}}_x}|{\bf {y}})}[(D_{\theta_{dx}}(G_{\varphi_x}({\mathbf{z}}_{x})))^2] \\
  & \quad + {\mathbb E_{{\bf {x}} \sim q_{data}({\bf{x}})}} [(D_{\theta_{dx}}({\bf{x}})-1)^2].
  \label{GAN_dc}
   \end{aligned}
\end{equation}
In summary, the training of the DCCRN-VAE SE system involves three stages, as illustrated in Fig.~\ref{fig:DCCRN_VAE}(a)(b). First, we utilize (\ref{ELBO}) to (\ref{si_sdr_loss}) to train C-VAE and N-VAE to obtain fully disentangled speech and noise representations. Following this, CC and DC encoders, along with CC and DC LSTM are frozen to train NS-VAE based on (\ref{latent_loss}). {The aim is to disentangle the clean speech representation from the observed signal}. Finally, the adversarial loss (\ref{GAN_dc}) is employed to train CC decoders, enhancing their generative ability. We can directly utilize NC encoders, NC LSTM, and CC decoders during SE when the observed signal is provided as input, as illustrated in Fig.~\ref{fig:DCCRN_VAE}(c) and red part in Fig.~\ref{fig:DCCRN_VAE}(b)).

\section{Experiment and result analysis}
\label{sec:experiment}

\begin{table}[!t]
 \centering
  \caption{{{Average evaluation score comparison on simulated dataset}}}
  \label{tab: simulated_score}
  \centering
    \begin{tabular}{cccccc}
    \toprule
    Method &SI-SDR &STOI (\%) &BAK &SIG &PMOS\\
    \midrule
    Noisy &11.12&81.07 &2.11  &2.46 & 2.86\\
    Oracle &49.92&99.97 & 4.06 & 3.57& 3.89\\
    NSNet2 &8.99&83.08 &3.78  &3.05 & 3.34\\
    DIS-VAE &11.03&84.02 &3.50  &3.08 & 3.38\\
    DCCRN & 16.43 & 87.01 & 3.77 & 3.19 &  3.47 \\
    DCCRN-VAE & 18.44 & 90.10  & 3.82 & 3.16  & 3.50 \\
    \bottomrule       
  \end{tabular}
\end{table}

\begin{table}[!t]
 \centering
  \caption{{{Average evaluation score comparison on real recordings}}}
  \label{tab: real_score}
  \centering
    \begin{tabular}{cccc}
    \toprule
    Method &BAK &SIG &PMOS\\
    \midrule
    Noisy & 2.34 & 2.88 & 2.91 \\
    NSNet2 & 3.80& 2.90& 3.29\\
    DIS-VAE & 3.53  & 3.01 & 3.33\\
    DCCRN & 3.68 & 3.05 & 3.39\\
    DCCRN-VAE & 3.81 & 3.07 & 3.43\\
    \bottomrule       
  \end{tabular}
    \vspace{-0.2cm}
\end{table}

This section will evaluate the DCCRN-VAE's SE performance. The aim is to assess the impact of the proposed DRL strategy on the DCCRN and DIS-VAE SE systems. To achieve this, we select several reference methods, including DCCRN \cite{hu20g_interspeech},  DIS-VAE \cite{xiang2022two}, and the DNS 2021 challenge baseline NSNet2 \cite{braun2020data, xia2020weighted}. The aim of comparing DCCRN-VAE with DIS-VAE is to determine whether a more sophisticated probabilistic model and DNN structure can yield improvements for DIS-VAE. The structures of DIS-VAE and DCCRN are described in \cite{xiang2022two} and \cite{hu20g_interspeech}, respectively. {Comparing DCCRN-VAE with DCCRN can help us understand whether a better DRL strategy can enhance the STOA DCCRN's SE performance. Additionally, NSNet2 is used as our SE baseline.}

{\bf Datasets:} This work evaluates the proposed algorithm using the DNS challenge 2021 corpus \cite{reddy2021interspeech}. All clean speech data are randomly divided into 70\% for training, 20\% for validation, and 10\% for evaluation. Similarly, the noise data from the DNS noise corpus are also split into training, validation, and test sets in proportions similar to those used for the speech data. Subsequently, the corresponding speech and noise corpora for training, validation, and testing are randomly mixed using the DNS script \cite{reddy2021interspeech}, with random signal-to-noise ratio (SNR) levels ranging from -10dB to 15dB. Other parameters for signal mixing follow the default values specified in the DNS script \cite{reddy2021interspeech}. Finally, a dataset is constructed, consisting of 500 hours of mixed training utterances, 10 hours of mixed validation utterances, and 3 hours of mixed test utterances. All signals are down-sampled to 16 kHz. Additionally, all real-world noisy recordings from the DNS challenge 2021 test dataset, which are collected from various complex noisy environments, are used for real evaluation purposes. Further data details can be found in \cite{reddy2021interspeech}.

{\bf Experimental setups:} In the proposed framework, all the VAE neural structures are designed to be similar to DCCRN \cite{hu20g_interspeech} for fair comparisons. Specifically, C-VAE and N-VAE share the same architecture with the following specifications: the number of channels is set to \{32, 64, 128, 128, 256, 256\}, the kernel size and stride are (5,2) and (2,1), respectively, and the complex LSTM has 128 units for both the real and imaginary parts. NS-VAE closely resembles C-VAE but excludes the decoder part. The discriminator structure is akin to NS-VAE but employs a real LSTM with 1 unit. During our experiments, the signal frame length was set to 25 ms with a frame shift of 6.25 ms, and the FFT length was 512. We utilized the Adam optimizer with a learning rate of 0.001 for training.

{\bf Experimental results:} To evaluate the SE performance of various algorithms, we will employ SI-SDR \cite{le2019sdr}, short-time objective intelligibility (STOI) \cite{taal2011algorithm}, and DNSMOS P.835 \cite{reddy2021dnsmos, reddy2022dnsmos}. DNSMOS P.835 enables us to assess the speech quality (SIG), background noise quality (BAK), and overall quality (PMOS) of the audio samples. DNSMOS P.835 has demonstrated a high degree of alignment with human ratings for evaluating speech quality. Importantly, DNSMOS P.835 does not require reference clean speech, making it suitable for evaluating the quality of real-world recordings.

First, we evaluate the SE performance of various algorithms on the simulated dataset. The results are presented in Table~\ref{tab: simulated_score}. 'Oracle' indicates the ultimate performance of the DCCRN-VAE system when NS-VAE and C-VAE share the same latent representation, where both the KL loss $\mathcal{L}_{kl}$ and the residual loss $\mathcal{L}_{re}$ in (\ref{latent_loss}) are set to 0. The 'Oracle' serves as an indicator of the method's potential. Comparing DCCRN-VAE to DCCRN, we observe significant improvements in SI-SDR, STOI, BAK, and PMOS. These results demonstrate that our DRL strategy enhances the SE performance of the current SOTA DNN-based SE algorithm. Furthermore, comparing DCCRN-VAE to DIS-VAE, we observe a notable enhancement across all five metrics. This confirms that a more sophisticated probabilistic model and DNN structure are crucial for achieving superior SE performance in DRL-based systems.

The experiments also include evaluations using real-world recordings, and the results are presented in Table~\ref{tab: real_score}. These results show that DCCRN-VAE exhibits a significant improvement in BAK compared to DCCRN and DIS-VAE. Additionally, DCCRN-VAE achieves higher scores in SIG and PMOS, which illustrates that the proposed DCCRN-VAE can effectively suppress background noise while simultaneously enhancing speech quality in practical applications.  {In summary, DCCRN-VAE leverages the advantages of the DRL strategy \cite{xiang2022two} and the DCCRN architecture, resulting in improved SE performance. Furthermore, DCCRN-VAE has the potential for further enhancement if more advanced methods are employed for DRL.}

\vspace{-10pt}
\section{Conclusion}

Building upon our preliminary DIS-VAE work \cite{xiang2022two}, this paper introduces a novel DCCRN-VAE SE system. Specifically, DCCRN-VAE incorporates a sophisticated DCCRN neural network structure and complex Gaussian distribution in DIS-VAE. {This strategy also improves the DCCRN's representation learning ability. Experimental results show that DCCRN-VAE can take advantage of DCCRN-VAE and DCCRN, achieving significant improvements in all five evaluation metrics (SI-SDR, STOI, BAK, SIG, and PMOS) for both simulated and real datasets.} Moreover, based on the results obtained with the oracle configuration, we believe that DCCRN-VAE's SE performance can be further enhanced by introducing more advanced DRL strategies to reduce the latent losses in (\ref{latent_loss}). 
This work solely utilizes the fundamental KLD and $L_2$-norm for latent loss calculation.

  
\ninept 

\scriptsize 
\bibliographystyle{IEEEtran}
\bibliography{IEEEabrv,myabrv_new,my_reference}
\end{document}